\documentclass [12pt,a4paper]{article}
\usepackage{amssymb,verbatim}            

\newtheorem{thm}{Theorem}[section]

\newtheorem{defn}[thm]{Definition}

\newtheorem{conc}[thm]{Conclusion}

\newtheorem{hypo}[thm]{Hypothesis}
\newtheorem{assump}[thm]{Assumption}

\newcommand{\R}{\mathbb R}

\newcommand{\RRR}{{\mathbb R}^3}
\newcommand{\RRRR}{{\mathbb R}^4}
\newcommand{\bQ}{\mathbb Q}

\title{\bf\large Physics and the Measurement of Continuous
Variables\thanks{To appear in \emph{Foundations of Physics}. The
published version will be available at www.springerlink.com, DOI
10.1007/s10701-007-9203-z}}

\author{{\normalsize R. N. Sen}\\
{\normalsize\em Department of Mathematics and Computer Science}\\
{\normalsize\em Ben-Gurion University, 84105 Beer Sheva, Israel}\\[2mm]
{\normalsize E-mail: rsen@cs.bgu.ac.il}\\ \quad}
\date{}

\vspace{-5mm}

\begin{document}

\maketitle
\thispagestyle{empty}

\begin{abstract}

This paper addresses the doubts voiced by Wigner about the physical
relevance of the concept of geometrical points by exploiting some
facts known to all but honoured by none: Almost all real numbers are
transcendental; the explicit representation of any one will require
an infinite amount of physical resources. An instrument devised to
measure a continuous real variable will need a continuum of internal
states to achieve perfect resolution.  Consequently, a laboratory
instrument for measuring a continuous variable in a finite time can
report only a finite number of values, each of which is constrained
to be a rational number. \emph{It does not matter whether the
variable is classical or quantum-mechanical.} Now, in von
Neumann's measurement theory \cite{JvN}, an operator $A$ with a
continuous spectrum -- which has no eigenvectors -- cannot be
measured, but it can be approximated by operators with discrete
spectra which are measurable.  The measurable approximant $F(A)$ is
not canonically determined; it has to be chosen by the
experimentalist.  It is argued that this operator can always be
chosen in such a way that Sewell's results \cite{SEW-05a,SEW-05b} on
the measurement of a hermitian operator on a finite-dimensional
vector space (described in Sec.\ \ref{FINITE-APPROACH}) constitute an
adequate resolution of the measurement problem in this theory.  From
this follows our major conclusion, which is that the notion of a
geometrical point is as meaningful in nonrelativistic quantum
mechanics as it is in classical physics. It is necessary to be
sensitive to the fact that \emph{there is a gap between theoretical
and experimental physics}, which reveals itself tellingly
as an error inherent in the measurement of a continuous variable.

\end{abstract}
\pagebreak

\hfill
\begin{minipage}[t]{7.5cm}
{\small\emph{Mathematical descriptions are necessarily more refined
than the physical operations that they purport to represent.}

\vspace{2mm}\hfill{\small G. L. Sewell}}
\end{minipage}

\section*{Introduction}

At almost every step, theoretical physics makes the assumption that
space, time and space-time are continuua -- that is, they are locally
homeomorphic with $\RRR,\, \R$ and $\RRRR$ respectively. If one sets
up a cartesian coordinate system on any of them and chooses a point
at random, that point will have, \emph{with probability one}, at
least one irrational coordinate. Yet all measurements recorded in the
laboratory are expressed in terms of rational numbers.  The set of
rational numbers $\mathbb Q$ is is generally considered discrete,
i.e., endowed with the discrete topology (in which one-point sets are
open).

It is the view of many eminent mathematicians that ``\emph{Bridging
the gap between the domains of discreteness and of continuity} is a
central, presumably even \emph{the} central problem of the
foundations of mathematics'' (\cite{FRA-BAR-LEV}, p.211); yet this
gap -- which is precisely the gap between experimental and
theore\-tical physics mentioned above -- does not seem to have
attracted attention in physics itself, despite the dependence of
physics on mathematics.

The problem reveals itself as soon as one considers infinitely
precise measurements of a continuous variable.  In this article we
shall analyse the problem, suggest a solution and briefly explore
some implications of the solution. 

The plan of this work is as follows. In {Sec.}~\ref{M-O-L} we shall
make precise the notion of a `perfect' classical instrument for
measuring lengths (the basic continuous variable in physics) in terms
of the resources it would demand, and formulate these requirements as
finiteness conditions.  In {Sec.}~\ref{M-S-O} we shall return to the
problem of single measurements in quantum mechanics that was
emphasized by Wigner \cite{WIG-52,WIG-81}, and reconsider his
conclusions in the light of von Neumann's original formulation and
the finiteness conditions of {Sec.}~\ref{M-O-L}.  In
{Sec.}~\ref{P-M-QM} we shall sketch the main problem of von Neumann's
theory of measurement in quantum mechanics, paying attention to
the measurement of operators with continuous spectra, which is often
neglected in the literature. After a brief consideration of the
results of Hepp \cite{HEP}, we shall proceed to the results recently
obtained by Sewell \cite{SEW-05a,SEW-05b}.  In {Sec.}~\ref{TWO-VER},
we shall suggest the division of measurement theory into two:
i)~$\R$-\emph{measurement theory}, in which no finiteness conditions
are imposed and the problems arising from continuous spectra are
addressed via the weakest possible hypothesis;
ii)~$\bQ$-\emph{measurement theory}, in which finiteness conditions
appropriate to laboratory physics are imposed, and a selection
criterion, again appropriate to laboratory physics, is proposed to
address the problems arising from continuous spectra.  It will be
seen that Sewell's work provides an adequate resolution of the
problem in $\bQ$-measurement theory. In {Sec.}~\ref{REMARKS} we shall
remark upon some possible implications.

\section{Measurement of Length}\label{M-O-L}

We begin by stating our basic assumptions (in the form of
definitions). 

\begin{defn}\label{PT-LINE}
In the following, the terms \emph{point} and \emph{line} will be
used in the sense of Euclidean geometry. The Euclidean line will be
assumed to be homeomorphic with the set of real numbers $\mathbb R$
with its usual topology \emph{\cite{SEN}}.
\end{defn}

\begin{defn}\label{PERFECT}
A length-measuring device will be said to be \emph{perfect} $($or to
have \emph{infinite resolution}$)$ if its states can be brought into
$(1,1)$ correspondence with points on the unit interval $[0,1]$. 
\end{defn}

We now consider the physical (memory) resources that would be
required to set up such a device.

Assume that we have a plane on which a line $l$ is marked.  Let $O$
and $B$ be fixed points on $l$, and $A$ any point on the segment
$OB$. We wish to measure the length $L(A)$ of the segment $OA$ with a
perfect measuring device, the scale of length being the one that is
built into the device.

A length measurement may be visualized as a two-step process, as
follows:

\begin{equation}\label{TWO-STEP}
A \mapsto \xi \mapsto  \alpha
\end{equation}

 \noindent At the first step, one maps the point $A$ to a state $\xi$
of the apparatus (for example, by aligning the point $A$ with the
cross-hairs).  At the second step, one carries out a read-write
operation, i.e., reads the numerical value $\alpha$ associated with
the state $\xi$ and writes it down. This numerical value will be the
length of the segment $OA$: $L(A) = \alpha$.

The reason for amalgamating the `read' and `write' operations into
one step is that we are assuming, following Bohr, that a measurement
is not completed until it is recorded.  However, the two can be
analysed separately.  A write operation will require a certain amount
of computer memory, paper and pencil, or whatever.

If $A$ is randomly chosen, then $L(A)$ will, with probability one, be
a transcendental number.\footnote{Recall that a real number is called
transcendental if it is not algebraic, i.e., is not the solution of
an algebraic equation. The set of all real algebraic numbers is
countable, and therefore of Lebesgue measure zero on the real line.
It follows that the Lebesgue measure of the set of transcendentals in
any interval equals the length of the interval.} In any digital
representation, a transcendental number has an infinite number of
digits. It will require an infinite amount of physical memory to
record it, and an infinitely large laboratory to hold the record.

Let us now turn to the read operation, that is the map $\xi \mapsto
\alpha$ in eq.\ (\ref{TWO-STEP}). This condition -- that the
measuring instrument return a numerical value without calling upon
devices external to it -- is required to eliminate the possibility of
infinite von Neumann chains

\begin{equation}\label{VON-NEUMANN-CHAIN}
A \mapsto \xi \mapsto \eta \mapsto \cdots \mapsto \alpha
\end{equation}

 \noindent To fulfil this condition the `reader' must be equipped with
a read-only memory cell for each state of the device, the cell for
the state $\xi_0$ containing the number $\alpha_0$ appropriate to it.
The number of memory cells required is the power of the continuum.
Therefore an infinite amount of physical memory will have to be built
into the reader.\footnote{Observe that the actual size of each memory
cell -- as long as it is nonzero -- is not relevant to the above
argument.} Again, an infinitely large laboratory will be needed to
accommodate it.

We may therefore articulate the following:

\begin{conc}\label{CONC-1} 
If the finiteness of physical resources is considered as a
constraint, the only physical quantities that can be measured
precisely are those that can assume only a finite number of values,
each possible value being a rational number.

\end{conc}

\noindent It should be emphasized that this conclusion derives
solely from the information content of real numbers.

An alternative statement of conclusion \ref{CONC-1} would be:

\begin{conc}\label{CONC-ALT} 
A $($digital\,$)$ measuring instrument can have only a finite number
of states $\xi$, and the measured value $\alpha_{\xi}$ associated
with the state $\xi$ must be a rational number.
\end{conc}

These conclusions hold both for classical and quantum-mechanical
systems. The concept of a perfect measurements is meaningful only if
one is willing to admit instruments that are infinitely large.

\section{Measurement of Single Observables\\ in Quantum Mechanics}
\label{M-S-O}

We now turn to considerations that are specific to quantum mechanics.
Our discussion will be based on the mathematically rigorous Hilbert
space formulation of quantum mechanics that was advanced by von
Neumann in his 1932 monograph \cite{JvN}. We shall therefore
understand the terms `measurement' and `the theory of measurement in
quantum mechanics' in the sense of von Neumann.  In this section we
shall review the measurement of single observables, preparatory to a
consideration of the main problem of von Neumann's theory in Sec.\
\ref{P-M-QM}.

von Neumann's development of the statistical interpretation of
quantum mechanics is based on the analysis of \emph{two successive
measurements} of a self-adjoint operator $A$ upon the same system,
the second measurement immediately following the first.\footnote{It
is the result of the second measurement that tells us whether the
system behaves in accordance with quantum mechanics, or the 1924
radiation theory of Bohr, Kramers and Slater (see \cite{JvN}, pp.\
213-214). This assumes that the result of an individual measurement
is error-free, or very nearly so. The nineteen-page paper by Bohr,
Kramers and Slater \cite{BKS} contains only one formula, $h\nu =
E_1-E_2$, and the modern reader will probably find it difficult to
read. It is therefore worth mentioning that the background to this
paper is explained in some detail in the Historical Introduction of
the volume \cite{BLvdW} edited by van der Waerden, in which this
paper is reprinted. Recall that in the Bohr-Kramers-Slater theory
energy and momentum are conserved statistically, but not in
individual events.} It is based on the assumption that
\emph{individual measurements can be performed}, and the first
question he addresses concerns the accuracy of these measurements. It
suffices to consider the following three cases\,\footnote{These
considerations may easily be extended to include operators with mixed
spectra.} (see \cite{JvN}, pp.\ 211-220): 

\begin{enumerate}

\item The spectrum of $A$ is discrete and nondegenerate. In this case
$A$ can be measured precisely. If the state before measurement is a
superposition of eigenvectors of $A$, then the state after
measurement is a unique eigenvector of $A$, which is determined by
the measured value. 

\item The spectrum of $A$ is discrete, but degenerate. In this case,
$A$ can be measured precisely, but the state after measurement ``is
not uniquely determined by the knowledge of the result of the
measurement'' (\cite{JvN}, p.\ 218). However, the second measurement
will return the same numerical value as the first.\footnote{If there
exists a self-adjoint operator $B$ which commutes with $A$ and lifts
the degeneracy, then $A$ and $B$ may be measured simultaneously. The
state after such a simultaneous measurement is again a unique joint
eigenvector of $A$ and $B$.} 

\item The spectrum of $A$ is continuous. In this case, $A$
\emph{cannot} be measured precisely.\footnote{This is, in some sense,
obvious; eigenvectors belonging to the continuous spectrum (such as
plane waves) do not lie in the Hilbert space.} However, $A$ can be
measured approximately, in the following sense. Let $\lambda_k, k \in
{\mathbb Z}$ be points on the real line such that $\lambda_k <
\lambda_{k+1}$. Then one can determine an interval $(\lambda_n,
\lambda_{n+1})$ in which $A$ lies. We may, without loss of
generality, assume that $\lambda_{k+1} - \lambda_k = \epsilon$ for
all $k$. Then $\epsilon$ is the measurement error. Finally, for each
$k \in {\mathbb Z}$, let $\lambda^k$ be any point in $(\lambda_k,
\lambda_{k+1})$. Then there exists an operator $F(A)$ with a discrete
spectrum that consists precisely of the points $\lambda^k, k \in
{\mathbb Z}$ such that an approximate measurement of $A$ is
equivalent to a precise measurement of $F(A)$ (\cite{JvN}, pp.\
220-221).\footnote{{} We shall call such an $F(A)$ a \emph{von
Neumann approximant} of $A$.}

\end{enumerate}

Thus von Neumann reduced approximate measurements of operators with
continuous spectra to exact measurements of operators with discrete
spectra. {{} This reduction was based on the mathematical results
that a self-adjoint operator $A$ with (only) a continuous spectrum
had no eigenvectors, but a plentiful supply of approximate
eigenvectors ($||A\phi - \lambda\phi||< \epsilon$).\footnote{This
fact forms the basis of the spectral theorem for bounded (and
eventually unbounded) self-adjoint operators; see, for example,
\cite{RIESZ-NAGY} or \cite{REED-SIMON}.}} However, from 1952, Wigner
began to express reservations about the very notion of exact
measurement in von Neumann's theory.  In his 1981 lecture notes he
wrote (\cite{WHE-ZUR}, p.\ 298):

\begin{quote} 

``Unfortunately, as we shall see, there are serious limitations on the
measurability of an arbitrary quantity. They blur the mathematical
elegance of von Neumann's original postulate that all self-adjoint
operators are measurable\footnote{Wigner's use of the word `all' in
this sentence is at odds with what he had written earlier in the
same notes. On p.\ 274 of \cite{WHE-ZUR}, he wrote: ``We now proceed
to the more general case in which $A$ may also have a continuous
spectrum. In this case one has to admit that measurement will not
yield a mathematically precise value -- no one asks whether the
outcome of this measurement is a rational or irrational
number.''}\ldots What then are the limitations of measurability? 

\emph{Only Quantities Which Commute with ALL Additive Conserved
Quantities Are Precisely Measurable.}''

\end {quote}

The first result in this direction was obtained by Wigner in 1952
\cite{WIG-52}. The conserved quantity was the $z$-component of the
angular momentum; the quantity measured was the $x$-component of the
spin of a spin-$\textstyle\frac12$ particle. In 1960, Araki and
Yanase proved that a bounded self-adjoint operator $\mathfrak{M}$
with discrete spectrum which does not commute with an additively
conserved quantity cannot be measured precisely \cite{ARA-YAN}. They
also showed that an approximate measurement could be carried out,
that is, the measurement error could be made smaller than any
$\epsilon > 0$, provided that the measuring apparatus was a
superposition of sufficiently many eigenstates, with different
eigenvalues, of the conserved quantity.\footnote{In 1961, Yanase
returned to the case considered by Wigner in 1952 and obtained a
lower bound for the error $\epsilon$ in terms of the ``size'' of the
apparatus, defined as the mean square of the conserved quantity over
the apparatus states \cite{YAN}. This result was attributed by Wigner
to Araki and Yanase in \cite{WHE-ZUR}, p.\ 304.}

Let us now consider, briefly, the measurement of position of a
point-particle not bound to a site. For such a particle the position
operators $\vec{X}$, if they exist, will surely have continuous spectra,
and will therefore be only approximately measurable. From von
Neumann's argument, one concludes that there will exist three
exactly-measurable operators ${\vec{X}}_{\Delta}$ with discrete
spectra such that the exact measurement of ${\vec{X}}_{\Delta}$ is
equivalent to a determination of the position of the particle within
a parallepiped of volume $\Delta$. 

In relativistic theories, position operators (such as the
Newton-Wigner position operators) may exhibit acausal behaviour.
Hegerfeldt has shown that, in a theory in which one-particle states
belong to irreducible representations of the inhomogeneous Lorentz
group with $m\geq 0$, the existence of such an operator contradicts
the principle of causality (\cite{HEG}; references to earlier works
may be found there). 

Wigner was greatly upset by this result. In \cite{WHE-ZUR}, p.\ 312
he wrote:

\begin{quote}

``No matter how one defines the position, one has to conclude that the
velocity, defined as the ratio of two subsequent position
measurements divided by the time interval between them, has a finite
probability of assuming an arbitrarily large value, exceeding $c$.
One either has to accept this, or deny the possibility of measuring
the position precisely, or even giving significance to this concept;
a very difficult choice!''

\end{quote}

However, it is also possible to view the facts\footnote{The facts
here are the problems of relativistic quantum theories, which are
necessarily field theories.} in a more conservative manner; namely,
that on a matter of such import, one should withhold judgment until
it has been determined whether or not nonrelativistic quantum
mechanics can stand as an autonomous, consistent physical theory. In
the latter endeavour, the most important ingredient that has been
missing is (or so the author believes) a resolution of the
measurement problem. 

{{} Assume that the measurement problem has been resolved.} Let now
$A, F(A)$ and $\epsilon$ be as earlier, and let $H$ be an additively
conserved quantity that does not commute with $A$. Assume further
that $F(A)$ is bounded, and that it does not commute with $H$. Then,
according to the results of Araki and Yanase, $F(A)$ can be measured
approximately, within an error $\epsilon$, for any $\epsilon > 0$.
{{} An approximate measurement of $F(A)$ is equally an approximate
measurement of $A$; the fact that $F(A)$ cannot be measured precisely
does not seem to change the situation qualitatively.  Whether
this qualitative picture stands up to quantitative scrutiny can only
be discussed in the context of a resolution of the measurement
problem. The problem, and its resolution by Sewell, are summarized
below.}

\section{The Problem of Measurement in Quantum Mechanics}
\label{P-M-QM}

As is well known, the answer offered by the von Neumann theory of
measurement raises a problem of considerable gravity:
Schr\"odinger's cat paradox.

In von Neumann's theory, the measuring device is regarded as an
assembly of microscopic quantum systems. There is no characterization
of the states of the device other than as vectors (or perhaps
subspaces) of a Hilbert space.  Begin by considering, as Wigner does
in \cite{WIG-63} and \cite{WIG-81}, the case when the state of the
system is an eigenstate of the operator being measured.

Let $\mathfrak H$ and $\mathfrak K$ be the Hilbert spaces of the
system and the apparatus respectively. Suppose that $A$ is the
operator being measured, the system has been prepared in an
eigenstate $\sigma^{(\nu)} \in \mathfrak H$ of $A$, and the initial
state of the apparatus is $a \in \mathfrak K$. The initial state of
the combined system is then $a\otimes\sigma^{(\nu)} \in {\mathfrak K}
\otimes {\mathfrak H}$. The interaction will change only the state of
the apparatus:

\begin{equation}\label{UNHELPFUL}
a\otimes\sigma^{(\nu)} \rightarrow a^{(\nu)}\otimes\sigma^{\nu}
\end{equation}

\noindent The above transition, unlike the reduction of the wave
packet, may be effected by a unitary time evolution.\footnote{This
equation is eq.\ (1) of Wigner's notes (\cite{WHE-ZUR},p.\ 328), and
also of his 1963 paper \cite{WIG-63}.} But all it does is to mirror
the state of one quantum-mechanical system by another
quantum-mechanical system. How is the state of the latter to be
determined? If the apparatus is an object of greater complexity than
the system, the problem of determining its state may be even more
complex. Coupling a second instrument to determine the state of the
first will be of little help if the second instrument is also a
quantum-mechanical system like the first.  

This example clearly shows that the `quantum-mechanical measurement
problem' consists of two distinct problems. The first is the problem
of individual measurements: if the system is known to be in an
eigenstate of a certain observable, how does the measurement reveal
which eigenstate it is in?  Without solving this problem, one cannot
distinguish, by experiment, between quantum mechanics and the
Bohr-Kramers-Slater theory. The theoretical problem of explaining the
reduction of the wave packet, that is, of explaining the result of
\emph{many} measurements upon identical copies of the system, becomes
well-posed only after the problem of individual measurements has been
resolved.

von Neumann's conscious ego hypothesis solves both of these problems.
It is, however, what mathematicians would call a very strong
hypothesis. One would like to obtain the same results under somewhat
weaker conditions.

\subsection{The infinite-system approach}

The quantum theory of systems with infinitely many degrees of freedom
was developed in the 1960's. It was based, not on Hilbert space but
on $C^*$- and $W^*$-algebras (see, for instance, \cite{SEW-02}). For
$N$-particle systems, the canonical commutation (and anticommutation)
relations had only one irreducible representation. But for a
countable infinity of degrees of freedom, there were many
inequivalent irreducible representations, as well as representations
that were reducible but not fully reducible (see \cite{SCH}, and
references cited therein).\footnote{Only one of these representations
contained a no-particle state, and could be handled by Fock space
methods.} Use of operator-algebraic methods established a measure of
control over these representations. It was shown by Lanford and
Ruelle that, for any $^{\star}$-representation $\pi$ of the algebra
of observables $\mathfrak{A}$, there existed an algebra that could be
interpreted as the algebra of observables measurable outside any
bounded region of space \cite{LAN-RUE}. They called these
\emph{observables at infinity}. Typically, these observables are
global (spatial) averages of local ones. (The observables at infinity
could all be multiples of the identity; if this was indeed the case, the
representation $\pi$ was said to have \emph{short-range correlations}.)

Hepp realized that individual measurements could be regarded as
determining the values of observables at infinity of infinite
quantum-mechanical systems. He then attempted to exploit this fact to
eliminate the conscious ego from von Neumann's measurement theory.
In 1972, using these observables as `pointers', he constructed a
scheme in which the interaction between the measuring device and the
observed system caused (i)~the pointer value to change, and (ii)~the
state vector of the observed system to collapse \cite{HEP}.  However,
these changes were only effected in the limit $t\rightarrow\infty$;
they could not be effected in finite times.

The reason is as follows. The pointer positions of the apparatus are
necessarily different in the initial and final states of the total
system (observed system plus apparatus). This implies that the
primary representations\footnote{Recall that a representation of the
algebra $\mathfrak{A}$ is called \emph{primary} if its centre is
trivial, and these include the irreducibles.} associated with the
initial and final states are unitarily inequivalent to each other.
For \emph{finite times}, time evolution is unitarily implemented in
the Hilbert space of the primary representation associated with the
initial state of the total system, which means that the pointer
remains in its null position.  A change of pointer position requires
a change of representation. While this could not be achieved in
finite time, Hepp succeeded in showing that it could be achieved in
the limit $t\rightarrow\infty$.

In 1975, Bell claimed that Hepp's results were not valid, because the
limit ``$t\rightarrow\infty$ never comes'' \cite{BEL}. Although his
argument was flawed,\footnote{It was based on a model for which the
Schr\"odinger picture does not exist.  This fact was also overlooked
by the editors of \cite{WHE-ZUR} (see their remarks on p.\ 782 of
\cite{WHE-ZUR}).} the point he raised deserves consideration.

Experimental physics is constrained by the finiteness of laboratory
size and available time. To be significant, (theoretical) results
that are obtained in the infinite volume and/or infinite time limit
must satisfy one further condition, namely the impossibility of
distinguishing, \emph{experimentally}, between the limiting state and
states at \emph{all} sufficiently large but finite $(t,
V)$.\footnote{In our opinion, this requirement as important as that
of mathematical rigour.} The requirement that time evolution be a
one-parameter group of automorphisms of the algebra supplies the
required stability, but does not provide an estimate of the rapidity
of convergence to the limit.

\subsection{The finite-system approach}
\label{FINITE-APPROACH}

The finite-system approach developed by Sewell is based on
schemes devised by van Kampen in 1954 \cite{NvK-54} (see also
\cite{NvK-62} and \cite{NvK-88}) and Emch in 1964 \cite{EMC} for
deriving the Pauli master equation (or a generalization of it). van
Kampen was more concerned with the physics of coarse-graining, i.e.,
with understanding how microscopic observables gave rise to
macroscopic ones, and how at the same time irreversibility arose from
the underlying time-reversible classical or quantum mechanics. Emch
was more concerned with finding a mathematically rigorous framework
that underlay the master equation. He wrote explicitly that the
``difficult problem\ldots{}of how to determine in a natural way [the
macroscopic observables] from an a priori given [set of microscopic
observables] will not be touched upon\ldots''.  Sewell's analysis
accepts the rigorous framework of Emch, but his macroscopic
observables are related to microscopic ones as envisaged by van
Kampen.

Sewell's synthesis of the van Kampen-Emch schemes may be described
as follows:

\begin{enumerate}
\item The instrument $\mathcal I$ is an object consisting of $N$
particles, governed by quantum mechanics. A pure quantum-mechanical
state of the instrument is a vector in an infinite-dimensional Hilbert
space $\mathfrak K$. Since $N$ is finite, von Neumann's uniqueness
theorem applies.

\item The full algebra $\mathcal B$ of observables of $\mathcal I$
contains an Abelian subalgebra $\mathcal M$. The elements of
$\mathcal M$ are macroscopic observables in the sense of van Kampen.

\item There are no superselection rules on $\mathfrak K$. (This
requirement is stated explicitly in Emch \cite{EMC}.) This means that
the centre of $\mathcal B$ consists of multiples of the identity. Put
differently, given any macroscopic observable $M \in {\mathcal M}$,
there is an observable $B \in {\mathcal B}\setminus{\mathcal M}$
which does not commute with $M$.\footnote{Had the macroscopic
observables commuted with every microscopic observable, the Hilbert
space $\mathfrak K$ would have split into superselection sectors, and
no observable, either microscopic or macroscopic, would have been
able to induce a transition from one sector to another.}

\item The spectra of the macroscopic observables $M \in {\mathcal M}$
are discrete. The Hilbert space $\mathfrak K$ decomposes into a set
of pairwise-orthogonal subspaces, each of which is the simultaneous
eigenspace of every observable in ${\mathcal M}$. These subspaces are
the quantum analogues of classical phase cells, and their
dimensionalities are astronomically large.  
\end{enumerate} 

Under these conditions, Sewell showed that the essential
conclusions of Hepp's analysis are reproducible on the laboratory
scale; that is, in a finite time, and with the microscopic system
${\mathcal S}$ coupled to an apparatus ${\mathcal I}$ of large but
finite size \cite{SEW-05a,SEW-05b}. $\mathcal I$ is as described
above, the Hilbert space of $\mathcal S$ is $\mathfrak H$ and
the Hamiltonians of $\mathcal S$ and $\mathcal I$ are $H$ and
$K$ respectively. The coupled system is conservative, its Hilbert
space is $\mathfrak{H} \otimes \mathfrak{K}$,\footnote{Sewell's
ordering of the factors $\mathfrak H$ and $\mathfrak K$ is the
opposite of Wigner's eq.\ (\ref{UNHELPFUL}) in \cite{WIG-63}.} and
its total Hamiltonian $H_c = H\otimes I_{\mathfrak K} + I_{\mathfrak
H}\otimes K +V$. Here $V$ is the interaction between $\mathcal S$ and
${\mathcal I}$, and $I_{\mathfrak H}, I_{\mathfrak K}$ the identity
operators on $\mathfrak H$ and $\mathfrak K$ respectively. Dynamics
of the coupled system is governed by the standard $N$-particle
Schr\"odinger equation.

Sewell assumes that the Hilbert space $\mathfrak H$ is
$n$-dimensional.\footnote{As will become clear later, this
assumption cannot be relaxed.} Then the eigenfunctions $u_r$ of $H$,
$Hu_r = \epsilon_ru_r, r = 1,\ldots,n$ form a complete orthonormal
set in $\mathfrak H$. The microscopic observables of $\mathcal S$ are
assumed to form an algebra $\mathcal A$ of bounded operators on
$\mathfrak H$.

Furthermore, $\mathcal M$ is assumed to consist of linear
combinations of a finite set of orthogonal projectors
$\{\Pi_{\alpha}|\alpha = 1, 2, \ldots,\nu\}$ that span ${\mathfrak
K}$. Then any element $M \in {\mathcal M}$ can be written in the form
$M = \sum_{\alpha=1}^{\nu} M_{\alpha} \Pi_{\alpha}$, where the
$M_{\alpha}$ are constants. The subspaces ${\mathfrak K}_{\alpha} =
\Pi_{\alpha}{\mathfrak K}$ of $\mathfrak K$ correspond to classical
phase cells.  Each such cell represents a macrostate of $\mathcal I$,
and is identified by the position of a pointer (or set of pointers)
in a measurement process. We shall call $\alpha$ the \emph{pointer
reading}.

The measuring instrument is so designed that the ${\mathcal S} -
{\mathcal I}$ coupling does not induce transitions between the
eigenstates ${u_r}$ of ${\mathcal S}$. When this holds, the
interaction $V$ will take the form
\begin{equation}\label{FORM-OF-V}
V = \sum_{r=1}^n P(u_r)\otimes V_r, 
\end{equation} 
where $P(u_r)$ is the projection operator for $u_r$ and the $V_r$ are
observables of $\mathcal I$. It follows from these that the
Hamiltonian of the coupled system has the form $H_c = \sum_{r=1}^n
P(u_r)\otimes K_r$, with $K_r = K + V_r + \epsilon_rI_{\mathfrak K}$.

The system $\mathcal S$ and apparatus $\mathcal I$ are prepared
separately in the initial states $\psi = \sum_rc_ru_r$ and $\Omega$
respectively, $\psi$ being a pure normalized state ($\sum_r |c_r|^2 =
1$) and $\Omega$ a density matrix. $\mathcal S$ and $\mathcal I$ are
coupled at $t=0$, so that the initial state of the coupled system is
$\Phi(0) = P(\psi) \otimes\Omega$. Its state at time $t > 0$ is then
given by $\Phi(t) = U^{\star}(t) \Phi(0) U(t)$, where $U(t) = \exp
(iH_ct)$.  Owing to the form of $H_c$, the state $\Phi(t)$ can be
written as
\begin{equation}\label{FORM-OF-PHI-T}
\Phi(t) = \sum_{r,s=1}^n \bar{c}_rc_sP_{r,s}\otimes \Omega_{r,s}(t)
\end{equation}
where $P_{r,s}$ is the operator in $\mathfrak H$ defined by $P_{r,s}f
= (u_s,f)u_r\;\forall\;f\in{\mathfrak H}$, and $\Omega_{r,s}(t) =
U^*_r(t)\,\Omega\, U_s(t)$, with $U_r(t)=\exp(iK_rt)$.

For $t > 0$, the time-dependent expectation value of the observable
$A\otimes M$ of the coupled system is, by definition, $E(A\otimes M)
= \mathrm{Tr}\, (\Phi(t)[A\otimes M])$. In particular,
$E(A)=E(A\otimes I_{\mathfrak K})$, and $w_{\alpha} = E(I_{\mathfrak
H}\otimes \Pi_{\alpha})$ is the probability that $\mathcal I$ is
found in the macrostate ${\mathfrak K}_{\alpha}$. Sewell further
showed that, owing to the Abelian character of $\mathcal M$, the
expectation functional $E$ is compatible with a unique conditional
expectation functional on $\mathcal A$ with respect to $\mathcal M$.
This functional has the form 
\begin{equation}\label{CONDITIONAL} 
E(A|{\mathcal M}) = \sum_{\alpha} \omega_{{\alpha}}(A)\Pi_{\alpha}. 
\end{equation} 
With this preparation, Sewell showed \cite{SEW-05b} that $E(A)$ and
$E(A|{\mathfrak K}_{\alpha})$ could be written as 

\begin{equation}\label{EXPECT}
E(A) = \sum\limits_{r=1}^n|c_r|^2(u_r, Au_r) + \sum_{r\neq s; r, s = 1}^n
       \sum\limits_{\alpha=1}^{\nu} F_{r,s;\alpha} \bar{c}_r c_s (u_r, Au_s)
\end{equation}
and (for $w_{\alpha}\neq 0$)
\begin{equation}\label{COND-EXPECT}
E(A|{\mathfrak K}_{\alpha}) =
\sum\limits_{r,s=1}^nF_{r,s;\alpha}\bar{c}_rc_s(u_r,Au_s)/w_{\alpha}, 
\end{equation}
where the coefficients $F_{r,s;\alpha}$ are defined by
\begin{equation}\label{DEF-F-R-S}
F_{r,s;\alpha} = \mathrm{Tr}\,(\Omega_{r,s}(t)\Pi_{\alpha}).
\end{equation}
They satisfy the following conditions: i)~$F_{r,s;\alpha} =
\bar{F}_{s,r;\alpha}$; ii)~$ 0 \leq F_{r,r;\alpha} \leq 1$; and
iii)~$\sum_{\alpha = 1}^n F_{r,r;\alpha} = 1$. It follows from these
that, for $z_1,\ldots,z_n \in {\mathbb C}$, the sesqui\-linear form
$\sum_{r,s=1}^n \bar{z}_rz_sF_{r,s;\alpha}$ is positive, from which
it follows that
\begin{equation}\label{POSITI}
F_{r,r;\alpha}F_{s,s;\alpha}\geq |F_{r,s;\alpha}|^2.
\end{equation}
The time evolution of the composite system is carried entirely by the
$F_{r,s;\alpha}$. 

With an instrument $\mathcal I$ designed for the purpose, a pointer
reading $\alpha$ should specify a unique microstate $u_r$ of
$\mathcal S$, and different microstates of $\mathcal S$ should give
different pointer readings; the map $\Gamma$ from the set of
microstates $r$ of $\mathcal S$ to the set of macrostates $\alpha$ of
$\mathcal I$ should be bijective (which requires $\nu = n$). When
this holds, the sum $\sum_{\alpha} F_{r,r;\alpha}$ reduces to the
single term $F_{r,r;\alpha(r)}$ (where $\alpha(r) = \Gamma(r)$), so
that $F_{r,r;\alpha(r)}= 1$ and $F_{s,s;\alpha(r)} = 0$ for $s \neq
r$. {{} It now follows from the positivity condition (\ref{POSITI})
that
\begin{equation}\label{DECOHERENCE}
F_{r,s;\alpha} = 0\;\,\mbox{\rm for}\;\, r\neq s.
\end{equation}
Then (\ref{EXPECT}) reduces to}
\begin{equation}\label{COLLAPSE}
E(A) = \sum\limits_{r=1}^n |c_r|^2 (u_r,Au_r),
\end{equation}
which shows that the wave packet has collapsed. Finally, setting
$A = I_{\mathfrak H}$ in (\ref{COND-EXPECT}) one finds that
$w_{\alpha(r)} = |c_r|^2$, so that (\ref{COND-EXPECT}) becomes
\begin{equation}\label{SECOND-MEAS}
E(A|{\mathfrak K}_{\alpha(r)}) = (u_r, Au_r)
\end{equation}
which shows that the state of $\mathcal S$ following the measurement
is the vector state $\psi = u_r$.

The interaction between the system $\mathcal S$ and the instrument
$\mathcal I$ must be such that the correspondence between microstates
of $\mathcal S$ and the macrostates of $\mathcal I$ -- namely the
bijective map $\Gamma$ -- is stabilized within a finite
interval $\tau$ which is just a microscopic observational
time.  This is realized in the finite Coleman-Hepp model studied by
Sewell \cite{SEW-05a}.

\subsection{The effect of $n$ upon the quality of $\mathcal I$}

The instruments $\mathcal I$ that we have been considering so far, in
which {eq.}~(\ref{DECOHERENCE}) is strictly valid, have been called
\emph{ideal} by Sewell; they do not admit of measurement errors. An
ideal instrument will be a useful analytical tool only if it can be
approximated sufficiently well in the laboratory. Sewell investigated
this question via a notion of \emph{normal} instruments, which took
the possiblity of measurement errors into account. When this was done, 
{eq.}~(\ref{DECOHERENCE}) was replaced by one that had an error term,
\begin{equation}\label{NORMAL-INSTR}
0 < 1 - F_{r,r;\alpha(r)} < \eta(N).
\end{equation}
The error $\eta(N)$ should be a strongly-decreasing function of $N$
which should tend to zero as $N\rightarrow\infty$.

In an exactly-soluble model that Sewell considered (the finite
Coleman-Hepp model), he found that $\eta(N) = \exp (-cN/n)$, where
$c$ is a positive constant of order unity and $n =
\dim\,\mathfrak{H}$.  As $n$ was small and fixed ($n=2$ in the
Coleman-Hepp model) and $N \sim 10^{24}$, the error term was utterly
negligible. However, if $n$ increased while $N$ was held fixed, the
arguments that led to the formula $\eta(N) = \exp (-cN/n)$ could no
longer be carried through. To get an idea of how the estimate for
$\eta$ might be affected, consider a measurement in which the pointer
reading represents the value of an intensive variable $v$ of
$\mathcal I$. In that case, each subspace ${\mathfrak K}_{\alpha}$
will carry a range $\sim n^{-1}$ of values of $v$, and under the
assumptions of the large deviation principle for the fluctuations of
macroscopic observables, a pointer reading corresponding to a typical
${\mathfrak K}_{\alpha}$ would carry a probability of the order of
$\exp (-cN/n^2)$ that the state of $\mathcal I$ did not lie in that
subspace. Thus $\mathcal I$ would not be a reliable instrument when
$n$ comes close to the order of $N^{\frac12}$.\footnote{I am
indebteded to the referee for pointing out the above.}

\subsection{Reconsideration of the measurement of continuous
variables}\label{RECON-MEAS-CONT-VARIABLES}

As pointed out by von Neumann, an observable $A$ that has only a
continuous spectrum cannot be measured precisely, because it has no
eigenvalues.  Write its spectral decomposition as 
$$ A = \int\limits^\infty_{-\infty} \lambda dE_{\lambda}.$$ 
Then any vector in any subspace ${\mathfrak H}_{\Delta\lambda} =
(E_{\lambda + \Delta\lambda} - E_{\lambda}) \cdot{\mathfrak H}$ of
the system Hilbert space ${\mathfrak H}$ is an approximate
eigenvector of $A$ near the spectral value $\lambda$.  This fact
makes it possible to define an infinity of operators with discrete
spectra, each of which can claim to be a measurable approximant to
the operator $A$.  Some further specification is needed to make the
notion of \emph{measurement of an operator with a continuous
spectrum} well-defined.

The least restrictive specification is clearly the following:

\begin{hypo}\label{ALL-F} Let $A$ be an operator with a continuous
spectrum. Then every von Neumann approximant $F(A)$ to $A$ should
be precisely measurable.
\end{hypo}

Accepting this hypothesis means, in Sewell's scheme, admitting every
possible value of $n$. In this case, as we noted above, Sewell's
results would not be applicable. 

The alternative to accepting hypothesis \ref{ALL-F} would be to
accept only a suitable subset of the von Neumann approximants $F(A)$.
The problem is that \emph{there is no canonical choice imposed by the
theory}. It would require an external agency to exercise the choice,
which would make it difficult to claim that quantum-mechanical
measurement theory emerges from quantum mechanics alone. This point
is worth repeating, and we present it as a conclusion:

\begin{conc}\label{CONT-SPEC-MEAS-CONC}
In the formulation of quantum mechanics on Hilbert space, a
self-adjoint operator $A$ with continuous spectrum does not have exact
eigenvalues, but only approximate ones. An approximate measurement of
$A$ is defined by von Neumann to be an exact measurement of $F(A)$,
an operator with discrete spectrum that approximates $A$. However, the 
approximant $F(A)$ is not uniquely defined by the theory, and has to
be chosen by an external agency. 
\end{conc}
Our further discussion will be based on the following assumption,
which is considerably weaker than von Neumann's conscious ego
hypothesis:

\begin{assump}\label{CHOICE-OF-F}
If the self-adjoint operator $A$ with a continuous spectrum is at all
measurable, its approximant $F(A)$ which is measured is determined 
by \emph{the design of the experiment}.
\end{assump}

This assumption takes cognizance of the following facts:

\begin{enumerate}

\item With few exceptions, it is virtually impossible to design, let
alone build, an instrument to measure a self-adjoint operator. 

\item A laboratory measurement of an unbounded operator such as the
momentum can only deal with a bounded subset of its spectrum. 

\item \emph{Increasing the resolution of the instrument is
usually accompanied by a reduction in the range of the variable
measured}. An instrument that is designed to determine the sixth or
seventh significant figure after the decimal point will almost surely
be designed under the assumption that the first three to five
significant figures are known. 

\end{enumerate}
In short, \emph{the experimentalist has considerable control on the
dimension $n$ of the Hilbert space $\mathfrak H$} in Sewell's scheme.
It is to be expected that this control will be exercised to ensure
the quality of $\mathcal I$, i.e., to keep $\exp\,(-cN/n^2)$
negligibly small. However, even a perfect measurement of $F(A)$ will
only mean that the error in the determination of $A$ is within the
bounds imposed by the definition of $F(A)$.

\section{Two Versions of Measurement Theory}
\label{TWO-VER}

In view of the above, it is our opinion that the subject called
measurement theory be divided into two in order to understand the
problem and to appreciate the significance of Sewell's results.  We
shall call these two $\mathbb R$-\emph{measurement theory} and
$\mathbb Q$-\emph{measurement theory} respectively.\footnote{In
mathematics, the symbols $\mathbb R$ and $\mathbb Q$ are used to
denote the sets of real and rational numbers respectively.}

\subsection{$\mathbb R$-measurement theory}
\label{R-M-T}

This will be the idealized version in which all finiteness conditions
-- on physical resources and on the time available to make a
measurement -- are ignored, {{} and hypothesis \ref{ALL-F} is
accepted}.  This is the framework that underlies the von
Neumann-Wigner theory of measurement in quantum mechanics. 
Owing to the hypothesis \ref{ALL-F}, Sewell's
results do not provide a resolution of the measurement problem in
this framework. 

$\R$-measurement theory can only be considered as an analytical tool.
One may perhaps be excused for thinking that, as an analytical tool,
it has yet to prove its usefulness.

\subsection{$\mathbb Q$-measurement theory}
\label{Q-M-T}

In this version of measurement theory, the finiteness of physical
resources (and of time available to the experimenter) is taken into
account, without sacrificing mathematical rigour. Quantum mechanics
is developed on Hilbert space, but it is accepted that an observable
that can be observed in the laboratory can assume only a finite
number of values, and these values have to be rational. For
continuous spectra, this means explicit acceptance of assumption
\ref{CHOICE-OF-F}. This version corresponds more closely to
experiment, but departs dramatically from theory; variables that are
continuous in the theory are measured as discrete in the laboratory,
with the inherent error that this implies. Put differently, the gap
between theoretical and experimental physics referred to earlier
reappears as the inherent error of measurement. In the opinion
of the present author, this should be considered to be a reassuring,
rather than a disturbing fact.

In this version of measurement theory, Sewell's results provide
an adequate resolution of the measurement problem in quantum
mechanics.

\section{Remarks}\label{REMARKS}

\begin{enumerate}

\item The `position operators' that can be measured in the
laboratory\footnote{To avoid ambiguity, we shall call such operators
$l$-\emph{measurable}.} are defined on finite-dimensional vector
spaces, and have the form $q_M =\sum{q_j E_j}$, where the sum is from
$j = 1$ to $j = J$, the $q_j$ are the values that can be returned by
the device, and $E_j$ are one-dimensional projection operators.
Clearly, $q^{\prime}_M = \sum{q^{\prime}_i E^{\prime}_i}$ is another
$l$-measurable position operator, and in general one will have $[q_M,
q^{\prime}_M] \neq 0$.  On the other hand, one may be able to define
an $l$-measurable momentum operator $p_M$ that commutes with $q_M$;
it is well known that the canonical commutation relations cannot be
realized on finite-dimensional vector spaces.
 
 \item Let $\{E_k|k = 1,\ldots{K}\}$ be a set of one-dimensional
projection operators on a $K$-dimensional vector space $V$ over the
complex numbers.  The algebra of these projection operators over the
real numbers will contain a set of $l$-measurable operators $p$ and
$q$, and therefore all $l$-measurable observables that can be
constructed from them. This algebra will clearly be abelian. However,
the superposition principle would not have been lost. 

\end{enumerate}

\section*{Acknowledgements}

The author would like to thank Professors N. Panchapakesan, H.  Reeh,
H. Roos and particularly G. L. Sewell for reading and criticizing
earlier versions of this paper. The errors that remain are, however,
his own.

\vspace{2cm}\noindent Dated 10 August 2007


\begin{thebibliography}{99}
\frenchspacing

\bibitem{ARA-YAN} Araki, H. and M. M. Yanase, Measurement of
quantum-mechanical operators, \emph{Phys. Rev.} {\bf 120}, 622
(1960).


\bibitem{BEL} Bell, J. S., On wave-packet reduction in the
Coleman-Hepp model, \emph{Helv. Phys. Acta} {\bf 48}, 93 (1975).

\bibitem{BKS} Bohr, N., H. A. Kramers and J. C. Slater, The quantum
theory of radiation, \emph{Phil. Mag.} {\bf 47}(281), 785 (1924).
Reprinted in \cite{BLvdW}. 


\bibitem{EMC}  Emch, G\'erard, Coarse-graining in Liouville space
and master equation, \emph{Helv.\ Physica Acta} {\bf 37}, 532
(1964). 

\bibitem{FRA-BAR-LEV} Fraenkel, A. A., Y. Bar-Hillel and A. Levy (in
collaboration with D. van Dalen), \emph{Foundations of Set Theory},
2nd revised edition, North-Holland, Amsterdam, 1973. Reprinted by
Elsevier, Amsterdam, 2001.

\bibitem{HEG} Hegerfeldt, G. C., Remark on causality and
particle localization, \emph{Phys. Rev.} {\bf D 10}, 3320 (1974).


\bibitem{HEP} Hepp, K., Quantum theory of measurement and
macroscopic observables, \emph{Helv. Phys. Acta} {\bf 45}, 237
(1972).


\bibitem{NvK-54} van Kampen, N., Quantum statistics of irreversible
processes, \emph{Physica} {\bf XX}, 603 (1954).

\bibitem{NvK-62} van Kampen, N., Fundamental problems in statistical
mechanics of irreversible processes, pp.\ 173-203 of
\emph{Fundamental Problems in Statistical Mechanics}, ed.\ E. D. G.
Cohen, North-Holland, Amsterdam, 1962.

\bibitem{NvK-88} van Kampen, N., Ten theorems about quantum
mechanical measurements, \emph{Physica} {\bf A 153}, 97 (1988).

\bibitem{LAN-RUE} Lanford, O. and D. Ruelle, Observables at infinity
and states with short-range correlations in statistical mechanics,
\emph{Commun. Math. Phys.} {\bf 13}, 194 (1969).

\bibitem{LON-BAU} London, F. and E. Bauer, \emph{La th\'eorie de
l'observation en m\'ecanique quantique}, Actualit\'es scientifiques
et industrielles: Expos\'es de physique g\'en\'erale, No. 775,
Hermann, Paris, 1939. English translation, including a new paragraph
by F. London, pp.\ 217-259 of \cite{WHE-ZUR}.


\bibitem{JvN} von Neumann, J., \emph{Mathematical Foundations of
Quantum Mechanics}, translated from the German by R. T. Beyer,
Princeton University Press, Princeton, 1955. (German original:
\emph{Mathematische Grundlagen der Quantenmechanik}, Verlag Julius
Springer, Berlin, 1932.)

\bibitem{REED-SIMON} Reed, M. and B. Simon, \emph{Methods of Modern
Mathematical Physics}, vol.\ I, Functional Analysis, Academic Press,
New York, 1972.

\bibitem{RIESZ-NAGY} Riesz, F. and B. Sz.-Nagy, \emph{Functional
Analysis}, Frederick Ungar, New York, 1955; translated from the
French by Leo F. Boron. (French Original: \emph{Le\c{c}ons d'analyse
fonctionelle}, 2nd ed., Akademiai Kiado, Budapest, 1953.)


\bibitem{SCH} Schweber, S. S., \emph{An Introduction to Relativistic
Quantum Field Theory}, Harper \&{} Row, New York, 1962.


\bibitem{SEN} Sen, R. N., Why is the Euclidean line the same as
the real line?, \emph{Foundations of Physics Letters}, {\bf 12}, 325
(1999).

\bibitem{SEW-02} Sewell, G. L., \emph{Quantum Mechanics and its 
Emergent Macrophysics}, Princeton University Press, Princeton, 2002.


\bibitem{SEW-05a} Sewell, G. L., On the mathematical structure
of quantum measurement theory, \emph{Rep. Math. Phys.} {\bf 56},
271 (2005).


\bibitem{SEW-05b} Sewell, G. L., Can the quantum measurement
problem be resolved within the framework of Schr\"odinger
dynamics?, Lecture given at the J. T. Lewis Memorial Conference,
Dublin, June 14-17, 2005.


\bibitem{BLvdW} van der Waerden, B. L., ed., \emph{Sources of Quantum
Mechanics}, North-Holland, Amsterdam, 1967; reprinted by Dover
Publications, New York, 2007.

\bibitem{WHE-ZUR} Wheeler, J. A. and W. H. Zurek, eds., \emph{Quantum
Theory and Measurement}, Princeton University Press, 1983.


\bibitem{WIG-52} Wigner, E. P., Die Messung quantenmechanischer
Operatoren, \emph{Z. Phys.} {\bf 133}, 101 (1952).


\bibitem{WIG-63}  Wigner, E. P., The problem of measurement,
\emph{Am. J. Phys.} {\bf 31}, 6 (1963), reprinted in
\cite{WIG-SR} and \cite{WHE-ZUR}.

\bibitem{WIG-SR}  Wigner, E. P., \emph{Symmetries and
Reflections}, The M.I.T. Press, Cambridge, MA and London, 1970.

\bibitem{WIG-81} Wigner, E. P., \emph{Interpretation of Quantum
Mechanics}, Lectures given in the Physics Department of Princeton
University during 1976, as revised for publication, 1981. Pp.\ 
260-314 of \cite{WHE-ZUR}. 

\bibitem{YAN} Yanase, M., Optimal measuring apparatus,
\emph{Phys. Rev.} {\bf 123}, 666 (1961).



\end{thebibliography}
\end{document}